\documentstyle[12pt]{article}
\begin{document}
\begin{center}
{\large \bf ELECTROMAGNETIC DETECTION OF AXIONS\\ }
\vspace{1cm}
{\bf Hoang Ngoc Long\footnote{Electronic address:hnlong@bohr.ac.vn}\\}
{\it Institute of
Theoretical Physics,
National Centre for Natural Science and Technology, \\
P.O.Box 429, Bo Ho,
Hanoi 10000, Vietnam}\\ \vspace{0.7cm}
{\bf Dang Van Soa\\}
{\it Department of Physics, Hanoi University of Mining and Geology,\\
Dong Ngac, Tu Liem, Hanoi, Vietnam.}\\
\vspace{0.5cm}
and\\
\vspace{0.5cm}
{\bf Tuan A. Tran\footnote{Electronic address: tatran@ictp.trieste.it}\\}
{\it International Centre for Theoretical Physics, Trieste, Italy}\\
\vspace{1cm}
{\bf Abstract\\}
\end{center}
\hspace{0.5cm}Photon-to-axion conversions
in the static electromagnetic fields are reconsidered in detail by using
the Feynman diagram techniques. The differential cross sections are
presented for the conversions in the presence of the
electric field of the flat condenser as well as in the magnetic field of the
solenoid. Based on our results a laboratory experiment for the production and
the detection of the axions is described. This experiment will exploit the
axion decay constant as well as the axion mass.\\
\begin {center}
\end{center}
\newpage
\hspace{0.5cm}In the 1970's, it was shown that the strong CP problem
can be solved~\cite{pq} by
 the introduction of a light pseudoscalar particle, called
the axion~\cite{w2}. However, no positive indication of its existence has
been obtained so far.
At present, the parameters of axions are constrained by laboratory
searches~\cite{mia}
and by astrophysical and cosmological considerations~\cite{turner}.

\hspace{0.5cm}A particle, if it has a two-photon vertex, may be
created by a photon entering an external electromagnetic (EM) field.
An axion is one of such particles.
Conversion of the axions into EM power in a resonant cavity was firstly
suggested by Sikivie~\cite{s}. He suggested that this method can be used
to detect the hypothetical galactic axion flux that would exist if
axions were the dark matter of the Universe. Various terrestrial
experiments to detect invisible axions by making use of their coupling
to photons have been proposed~\cite{kvb,sde,fhoo}, and the first result of such
experiments appeared recently~\cite{cha}.

\hspace{0.5cm}Recently, a photon regeneration experiment, using RF
photons,  was
described~\cite{hoo}. That experiment consists of two cavities which are
placed a small distance apart. A more or less homogeneous magnetic field
exists in both cavities. The first, or emitting is excited by incoming
RF radiation. Depending on the axion-photon coupling constant, a
certain amount of RF energy will be deposited in the second, or receiving
cavity. In Ref.~\cite{hoo} the author considered the problem by using the
classical method. By applying the Feynman diagram techniques we have
considered the conversion of the photons into gravitons in the
static~\cite{lst1} and periodic~\cite{lst2} EM fields. In this paper we
also apply this method to reconsider the EM conversion of the axions
in both the electric and magnetic fields.

\hspace{0.5cm}The axion mass and its couplings to ordinary particles
are all inversely proportional to the magnitude $v$ of the vacuum expectation
value that spontaneously breaks the $U_{PQ}(1)$ quasisymmetry which was
postulated by Peccei and Quinn and of which the axion is the
pseudo-Nambu-Goldstone boson. For the
axion-photon system a suitable Lagrangian density is given
by~\cite{turner,s}:
\begin{equation}
L = -\frac{1}{4}F_{\mu\nu}F^{\mu\nu} + g_{\gamma}\frac{\alpha}{4\pi}
\frac{\phi_a}{f_a}F_{\mu\nu}\tilde{F}^{\mu\nu} +
\frac{1}{2}\partial_{\mu}\phi_a
\partial^{\mu}\phi_a - \frac{1}{2} m_a^2 \phi_a^2[1+0(\phi_a^2/v^2)]
\label{1}
\end{equation}
where $\phi_a$ is the axion field, $m_a$ its mass, $\tilde{F}_{\mu\nu}=
\frac{1}{2}\varepsilon_{\mu\nu\rho\sigma}F^{\rho\sigma}$, and $f_a$
is the axion decay constant and is defined in terms  of the axion mass
$m_a$ by~\cite{s,cha}:
$f_a = f_{\pi}m_{\pi}\sqrt{m_u m_d}[m_a(m_u + m_d)]^{-1}$.
The coupling constant in~(\ref{1}) is model
dependent. Interaction of the axions to the photons arises from the triangle
loop diagram, in which two vertices are interactions of the photon to
electrically charged fermion and  an another vertex is coupling of the axion
with fermion. This coupling is model dependent and is given:
\[
g_{\gamma}=\frac{1}{2}
\left(\frac{N_e}{N}-\frac{5}{3}-\frac{m_d-m_u}{m_d+m_u}\right),
\]
where $N=\mbox{Tr}(Q_{PQ}Q^2_{color})$ and
$ N_e=\mbox{Tr}(Q_{PQ}Q^2_{em})$. Tr represents the sum over all
left-handed Weyl fermions. $Q_{PQ}$, $Q_{em}$, and
$Q_{color}$ are respectively the Peccei-Quinn charge, the electric charge,
and one of the generators of $SU(3)_c$. In this paper we are considering very
light, and very weak interacting invisible axions~\cite{dfsz}, however our
calculation is also valid for the heavy axions case. In
this model - the Dine-Fischler-Srednicki-Zhitnitskii (DFSZ) model, one has
$N_e=\frac{8}{3}N$ hence
the coupling constant  $g_{\gamma}(DFSZ)\simeq 0.36$.  In other model as
given by the
Kim-Shifman-Vainshtein-Zakharov~\cite{ksvz}, where the axions do not
couple to light quarks and leptons (hadronic axion), one
has $N_e=0$ and hence $g_{\gamma}(KSVZ)\simeq -0.97$.

\hspace{0.5cm}Consider the conversion
of the photon $\gamma$ with momentum q into the axion
$a$ with momentum p in an external electromagnetic field.

\hspace{0.5cm}For the abovementioned process, the relevant coupling is
the second term in~(\ref{1}).
Using the Feynman rules we get the following expression for
the matrix element
\begin{equation}
\langle p| M |q\rangle = -
\frac{g_{a\gamma}}{2(2\pi)^2\sqrt{q_0p_0}}\varepsilon
_{\mu}(\vec{q},\sigma)\varepsilon^{\mu\nu\alpha\beta}q_{\nu}
\int_{V} e^{i\vec{k}\vec{r}}F_{\alpha\beta}
^{class}d\vec{r}
\label{2}
\end{equation}
where $ \vec{k}\equiv \vec{q}-\vec{p}$ the momentum transfer to the EM field,
 $g_{a\gamma}\equiv g_{\gamma}\frac{\alpha}{\pi f_a}$=
$g_{\gamma}\alpha m_a(m_u+m_d)(\pi f_{\pi}m_{\pi}\sqrt{m_um_d})^{-1}$
and $\varepsilon^{\mu}(\vec{q},\sigma)$
represents the polarization vector of the photon.

\hspace{0.5cm}Expression~(\ref{2}) is valid for an arbitrary external
EM field. In the following we shall use it for two cases,
namely conversion in the electric field of a flat condenser and in the
static magnetic field of the solenoid. Here we use the following
notations: $q \equiv |
\vec{q} |,  p \equiv |\vec{p}|=(q_o^2-m_a^2)^{1/2}$
and $\theta$ is the angle between
$\vec{p}$ and $\vec{q}$.

\hspace{0.5cm}{\it Conversion in the presence of an electric field}.-- Now we
take the EM field as a homogeneous electric field of a flat condenser of
size $a\times b\times c$. We shall use the coordinate system with
the x axis parallel to the direction of the field, i.e.,
$F^{10}=-F^{01} = E.$ Then the matrix element is given by
\begin{equation}
\langle p| M^e |q\rangle = \frac{g_{a\gamma}}{(2\pi)^2\sqrt{q_0p_0}}
\varepsilon_{\mu}
(\vec{q},\sigma)\varepsilon^{\mu\nu 0 1}q_{\nu} F_e(\vec{k}),
\label{3}
\end{equation}
where a form factor for the electric region~\cite{s,kvb}
\[F_e(\vec{k}) = \int_{V} e^{i\vec{k}\vec{r}}E(\vec{r})
d\vec{r}.\]
The superscript $e$ in $M^e$ refers to the process taking place in the
presence of an electric field.\\
For a homogeneous field of intensity E we
have~\cite{lst1} \begin{equation}
F_e(\vec{k}) = 8E\sin(\frac{1}{2}ak_x)\sin(\frac{1}{2}b
k_{y})\sin(\frac{1}{2}ck_{z})(k_xk_yk_z)^{-1}.
\label{4}
\end{equation}

\hspace{0.5cm}Substituting~(\ref{4}) into~(\ref{3}) we find finally the
differential cross
section (DCS) of the conversion of the axions in the electric field of a
flat condenser of size $ a\times b\times c$
\begin{equation}
\frac{d\sigma^e(\gamma \rightarrow a)}
{d\Omega}=\frac{g^2_{a\gamma}E^2}{2(2\pi)^2}
\left[\frac{\sin(\frac{1}{2}a
k_x)\sin(\frac{1}{2}bk_y)
\sin(\frac{1}{2}ck_z)}
{k_xk_yk_z}\right]^{2}(q^2_y + q^2_z).
\label{5}
\end{equation}
{}From~(\ref{5}) we see that if the photon moves in the direction of the
electric
field i.e., $q^{\mu} = (q, q, 0, 0)$ then DCS vanishes. If the momentum
of photon is parallel to the  y axis, i.e.,
$q^{\mu}$ =( q, 0, q, 0 ) then Eq.~(\ref{5}) becomes:
\begin{eqnarray}
\frac{d\sigma^e(\gamma \rightarrow a)}
{d\Omega"}&=&\frac{32 g^2_{a\gamma}E^2q^2}{(2\pi)^2}
\left[\sin\left( \frac{ap\sin \theta \sin \varphi"}{2}\right)
\sin\left(\frac{b}{2}(q-p\cos \theta)\right)\right.\nonumber\\
          &\times&\left.\sin\left(\frac{cp\sin \theta\cos\varphi"}{2}
\right)\right]^2(p^2\sin^2\theta\sin\varphi"\cos\varphi"
(q-p\cos\theta))^{-2}.
\label{6}
\end{eqnarray}
where $\varphi"$ is the angle between the z axis and the projection of
$\vec{p}$ on the xz plane~\cite{lst2}.\\
\hspace{1cm}From~(\ref{6}) we have
\begin{equation}
\frac{d\sigma^e(\gamma \rightarrow a)}{d\Omega"}=\frac{2
g^2_{a\gamma}E^2a^2c^2}{(2\pi)^2q^2\left(1-\sqrt{1-\frac{m^2_a}{q^2}}\right)^2}
\sin^2\left[\frac{qb}{2}\left(1-\sqrt{1-\frac{m^2_a}{q^2}}\right)\right]
\label{7}
\end{equation}
for $\theta\approx 0$ and
\begin{equation}
\frac{d\sigma^e(\gamma \rightarrow a)}{d\Omega"} = \frac{8
g^2_{a\gamma}a^2E^2}{(2\pi)^2(q^2-m^2_a)} \sin^2\left(\frac{bq}{2}\right)
\sin^2\left(\frac{cq}{2}\sqrt{1-\frac{m^2_a}{q^2}}\right)
\label{8}
\end{equation}
for $\theta=\frac{\pi}{2}, \varphi"=0$.

\hspace{0.5cm}In the limit $m^2_a\rightarrow 0$, Eqs.~(\ref{7})
and~(\ref{8}) become, respectively,
\begin{equation}
\frac{d\sigma^e(\gamma\rightarrow a)}{d\Omega"}=\frac{g^2_{a\gamma}
q^2V^2E^2}{2(2\pi)^2} + O(m_a^4)
\label{9}
\end{equation}
and
\begin{equation}
\frac{d\sigma^e(\gamma \rightarrow a)}{d\Omega"} = \frac{8
g^2_{a\gamma}a^2E^2}{(2\pi)^2q^2} \sin^2\left(\frac{bq}{2}\right)
\sin^2\left(\frac{cq}{2}\right) + O(m_a^4) .
\label{10}
\end{equation}
\hspace{0.5cm}From~(\ref{9}) we see that DCS in
the direction of the axion motion depends quadratically on
the intensity E, the {\it volume V} of condenser,
and {\it the photon momentum} q.

\hspace{0.5cm}For  $V=1m\times 1m\times 1m$, the
intensity of the electric field  $E = \frac{100 kV}{m}$, the photon
length $\lambda = 10^{-5}$ cm, and $m_a\simeq 10^{-5} eV$~\cite{cha}
then in the DFSZ model the cross section  given by~(\ref{9}) is
$\frac{d\sigma(\gamma\rightarrow a)}{d\Omega}
\simeq 8\times 10^{-22}cm^2$, while by~(\ref{10})
$\frac{d\sigma(\gamma\rightarrow a)}{d\Omega}
\simeq 8\times 10^{-51}cm^2$. We see that the axion is mainly
created in the
direction of photon motion. This coincides with the one dimensional
solution which is basics for the experimental setups in Ref.~\cite{kvb}.
To obtain the results in the KSVZ model we only need to
note that $g^2_{\gamma}(KSVZ)\simeq 7.26
\times g^2_{\gamma}(DFSZ)$ .

\hspace{0.5cm}For the case in which $q^2\rightarrow m^2_a$,
Eqs.~(\ref{7}) and ~(\ref{8}) become
\begin{equation}
\frac{d\sigma^e(\gamma \rightarrow a)}{d\Omega} = \frac{2
g^2_{a\gamma}a^2c^2E^2}{(2\pi)^2} \sin^2\left(\frac{bq}{2}\right).
\label{71}
\end{equation}
\hspace{0.5cm}For  $a=c=1 m$, the
intensity of the electric field  $E = \frac{100 kV}{m}$,
and $m_a=1 eV$~\cite{cha}
then  the cross section is given by~(\ref{71}):
$\frac{d\sigma(\gamma\rightarrow a)}{d\Omega}
\simeq 8.1\times10^{-26}cm^2$

\hspace{0.5cm}{\it Conversion in the presence of a magnetic field}.-- Now we
consider the conversion in
a homogeneous magnetic field of the solenoid with a radius R and
a length h, and without loss of generality suppose that
direction of the magnetic field is parallel to the z axis, i.e.,
$F^{12}=-F^{21}=B.$
After some manipulations we get
\begin{equation}
\frac{d\sigma^m(\gamma\rightarrow a)}
{d\Omega}=\frac{g^2_{a\gamma}F_m^2(\vec{q}-\vec{p})}{2(2\pi)^2}
q^2\left(1 - \frac{q^2_z}{q^2}\right)
\label{11}
\end{equation}
where $F_m$ is a form factor for the magnetic region~\cite{s,lst1}:
\begin{equation}
F_m(\vec{k})=\frac{4\pi BR}{k_z\sqrt{k_x^2+k_y^2}}J_1(R
\sqrt{k_x^2+k_y^2})\sin\left(\frac{hk_z}{2}\right).
\label{12}
\end{equation}
where $J_{1}$ is the one-order spherical Bessel function
(for a homogeneous magnetic field of magnitude B with a
size $a\times b\times c$ one has formula~(\ref{4}) with replacement E by B).

\hspace{0.5cm}From~(\ref{11}) it follows that when the momentum of the
photon is parallel to the z
axis (the direction of the magnetic field), DCS
vanishes. It implies that {\it if the momentum of the photon is parallel to
the EM field then there is no conversion}.
If the momentum of the
photon is parallel to the x axis, i.e., $q^{\mu}=(q,q,0,0)$ then
Eq.~(\ref{11}) gets the final form:
\begin{eqnarray}
\frac{d\sigma^m(\gamma\rightarrow a)}
{d\Omega'}&=&2g^2_{a\gamma}R^2
B^2J_1^2\left(Rq\sqrt{\left(1-\cos\theta\sqrt{1 - \frac{m_a^2}{q^2}}\right)^2 +
\left(1-\frac{m_a^2}{q^2}\right)\sin^2\theta\cos^2\varphi'}\right)\nonumber\\
          &\times&\left[\left(1-\cos\theta\sqrt{1-\frac{m_a^2}{q^2}}\right)^2+
\left(1-\frac{m_a^2}{q^2}\right)\sin^2\theta\cos^2\varphi '
\right]^{-1}q^{-2}  \nonumber \\
          &\times&\sin^2\left(\frac{hq}{2}\sqrt{1 -
\frac{m_a^2}{q^2}}\sin\theta\sin\varphi'\right)\left[(1-\frac{m_a^2}{q^2})
\sin^2\theta\sin^2\varphi'^2\right]^{-1}
\label{13}
\end{eqnarray}
where $\varphi'$ is the angle between the y axis and the projection of
$\vec{p}$ on the yz plane~\cite{lst1}.\\
It is easy to see that
\begin{equation}
\frac{d\sigma^m(\gamma\rightarrow a)}
{d\Omega'}=\frac{1}{2}g^2_{a\gamma}R^2h^2
B^2 J_1^2\left[Rq\left(1 - \sqrt{1 - \frac{m_a^2}{q^2}}\right)\right]
\left(1 - \sqrt{1 - \frac{m_a^2}{q^2}}\right)^{-2}
\label{14}
\end{equation}
for $\theta\approx 0$ and
\begin{equation}
\frac{d\sigma^m(\gamma\rightarrow a)}
{d\Omega'}=\frac{1}{2}g^2_{a\gamma}R^2h^2
B^2 J_1^2\left(Rq\sqrt{2 - \frac{m_a^2}{q^2}}\right)
\left(2 -  \frac{m_a^2}{q^2}\right)^{-1}
\label{15}
\end{equation}
for $\theta = \frac{\pi}{2}, \varphi'=0$.\\
For the limit $m^2_a\ll q^2$ with the notice that
\[\lim_{p\to q}\frac{J_{1}\left(R(q-p)\right)}{q-p}=\frac{R}{2}\]
then Eqs.~(\ref{14}) and~(\ref{15}) become, respectively,
\begin{equation}
\frac{d\sigma^m(\gamma\rightarrow a)}
{d\Omega'}=\frac{g^2_{a\gamma}V^2B^2q^2}{2(2\pi)^2} + O(m_a^4), \hspace{1cm}
V\equiv \pi R^2 h
\label{16}
\end{equation}
and
\begin{equation}
\frac{d\sigma^m(\gamma\rightarrow a)}
{d\Omega'}=\frac{1}{4}g^2_{a\gamma}R^2h^2B^2
J_1^2\left(\sqrt{2} R q\right) + O(m_a^4).
\label{17}
\end{equation}
\hspace{0.5cm}From~(\ref{16}) we see that DCS in
the direction of photon motion depends quadratically on
the magnitude B, {\it the cavity volume V} and {\it the photon momentum q}.

\hspace{0.5cm}From~(\ref{17}) it follows that DCS
vanishes when $p_n=\frac{\mu_n}{R\sqrt{2}}$ with
$n=0,\pm1\pm2$...and has its largest value
\begin{equation}
\frac{d\sigma^m(\gamma\rightarrow a)}
{d\Omega'}=\frac{1}{4}g^2_{a\gamma}R^2h^2B^2J_1^2(\mu'_{n})
\label{18}
\end{equation}
for $p_n=\frac{\mu'_n}{R\sqrt{2}}$, where $\mu_n$ and $\mu'_n$
are the roots of $J_1(\mu_n)=0$ and $J'_1(\mu'_n)=0$.

\hspace{0.5cm}For the magnetude~\cite{cha}  B = 8T,$ R = h = 1 m$,
$\lambda=10^{-5}cm$ and $m_a\simeq 10^{-5}eV$ by  (\ref{16}) we have
$\frac{d\sigma(\gamma\rightarrow a)}{d\Omega}= 4.7\times 10^{-11} cm^2$.

\hspace{0.5cm}For the limit $q^2\rightarrow m^2_a$, Eqs.~(\ref{14})
and~(\ref{15}) become:
\begin{equation}
\frac{d\sigma^m(\gamma\rightarrow a)}
{d\Omega'}=\frac{1}{2}g^2_{a\gamma}R^2h^2B^2
J_1^2\left(Rq\right).
\label{20}
\end{equation}
\hspace{0.5cm}For B = 8 T, $m_a = 1 eV$ and $R = h = 1 m$, we have
$\frac{d\sigma_(\gamma\rightarrow a)}{d\Omega}= 1.7\times 10^{-16} cm^2$.

\hspace{0.5cm}In the case of the magnetic field of size $a\times b\times
c$ we get a formula similar to (\ref{16}) under the same conditions.

\hspace{0.5cm}It is easy to show that the cross section for the reverse
process coincides exactly with above results, so that for conversion
photon-axion-photon, cross section is square of the previous
evaluation.

\hspace{0.5cm}Based on the above we describe an experiment:
The initial photon of energy $q_o$ from the laser (maybe better from X ray)
interacts with a virtual photon from the EM field to produce
the axion of energy $q_o$ and momentum $p=(q_o^2-m_a^2)^{1/2}$.
The photon beam is then blocked to eliminate everything except the
axions, which penetrate the wall because of their extremely weak
interaction with ordinary matter. (Such shielding is straightforward
for a low-energy laser beam.) The axion then interacts with another
virtual photon in the second EM field to produce a real photon of energy
$q_o$, whose detection is the signal for the production of the axion.
For the details of experimental setup the reader can see
Refs.~\cite{kvb,3a}.

\hspace{0.5cm}We note again that here EM field is understood by not only
magnetic field but also {\it the electric field of the condenser}.
To differ the axion signal from noise the case in which the EM field is
switched off
has to be measured. From the results we can get limits for $f_a$ as
well as on the
axion mass. The axion-photon coupling constant in~\cite{hoo} is taken
$g = 10^{-8}~\mbox{GeV}^{-1}$, this coupling constant corresponds to the
axion mass $m_a\sim 0.1~\mbox{eV}$.

We emphasize that the cross
section depends quadratically on the momentum of incoming photons hence high
frequencies are preferred despite the technical difficulties discussed
in~\cite{hoo}.

\hspace{0.5cm}In conclusion, the axion hypothesis can be tested experimentally.
Relatively simple experiments
can provide new information about physics at very high energies.
If the axion exists, we will have new powerful tools to study
the galaxy and the sun.

\hspace{0.5cm} We thank Professor J. Tran Thanh Van for
help and partial support.
T. A. T. would like to thank International Centre for Theoretical Physics for
financial support. D.V.S thanks Prof. N. S. Han for encouragement.\\

\end{document}